\documentclass[a4paper,fleqn,twocolumn]{cas-dc}

\usepackage[numbers]{natbib}

\usepackage{keyval,amsfonts,slashed,bm}
\usepackage{graphicx,textcomp}
\usepackage{epsfig,amsmath,amssymb}
\usepackage{subfig}
\usepackage{caption}
\usepackage{color}

\newcommand{\ba}{\begin{eqnarray}}
	\newcommand{\ea}{\end{eqnarray}}

\definecolor{sinopia}{rgb}{0.8,0.25,0.04}

\definecolor{greenopia}{rgb}{0.3,0.65,0.14}

\DeclareMathOperator*{\SumInt}{%
	\mathchoice%
	{\ooalign{$\displaystyle\sum$\cr\hidewidth$\displaystyle\int$\hidewidth\cr}}
	{\ooalign{\raisebox{.14\height}{\scalebox{.7}{$\textstyle\sum$}}\cr\hidewidth$\textstyle\int$\hidewidth\cr}}
	{\ooalign{\raisebox{.2\height}{\scalebox{.6}{$\scriptstyle\sum$}}\cr$\scriptstyle\int$\cr}}
	{\ooalign{\raisebox{.2\height}{\scalebox{.6}{$\scriptstyle\sum$}}\cr$\scriptstyle\int$\cr}}
}

\ExplSyntaxOn
\keys_set:nn { stm / mktitle } { nologo }
\ExplSyntaxOff

\begin{document}
\let\WriteBookmarks\relax
\def\floatpagepagefraction{1}
\def\textpagefraction{.001}
\shorttitle{QCD mesonic screening masses using Gribov quantization}
\shortauthors{Sumit et~al.}

\title [mode = title]{QCD mesonic screening masses using Gribov quantization}                      

\author{Sumit}[orcid=0000-0001-7137-6433]
\ead{sumit@ph.iitr.ac.in}
\address{Department of Physics, Indian Institute of Technology Roorkee, Roorkee 247667, India}
\author{Najmul Haque}[orcid=0000-0001-6448-089X]
\ead{nhaque@niser.ac.in}
\address{School of Physical Sciences, National Institute of Science Education and Research, HBNI, Jatni 752050, India}
\author{Binoy Krishna Patra}
\ead{binoy@ph.iitr.ac.in}
\address{Department of Physics,
	Indian Institute of Technology Roorkee, Roorkee 247667, India}


\begin{abstract}
The screening masses of mesons provide a gauge invariant and definite order parameter of chiral symmetry restoration. Different mesonic correlation lengths for flavor non-singlets, at least up to NLO, are well-defined gauge invariant physical quantities calculated earlier using the perturbative resummation techniques.
The NLO perturbative results match the available non-perturbative lattice QCD results at the high-temperature regime. 
We have studied the spatial correlation lengths of various mesonic observables using the non-perturbative Gribov resummation, both for quenched QCD and $(2+1)$ flavor QCD. 
The study follows the analogies with the NRQCD effective theory, a well-known theory for studying heavy quarkonia at zero temperature. 
\end{abstract}




\maketitle

\bigskip

\section{Introduction}\label{Int}	
Understanding experimental results at the relativistic heavy-ion colliders in high-energy nuclear and particle physics and the early cosmos' cosmic history depends heavily on quantum chromodynamics (QCD) at finite temperatures. When the temperature rises, the theory switches from a confined phase with hadronic degrees of freedom, where chiral symmetry is spontaneously broken, to a deconfined phase of quarks and gluons, where chiral symmetry is restored.

The temporal and spatial directions are generally disconnected because the heat bath destroys the Lorentz symmetry at finite temperatures. The temporal correlation functions in the Minkowski space can be used to define spectral functions for various operators in Fourier space. The spectral functions then indicate the plasma's essential ``real-time" features, such as the particle production rates~\cite{McLerran:1984ay}. In investigations involving heavy ion collisions, these are then immediately observable.	Nevertheless, the correlation functions in the spatial direction answer questions like At what length scales thermal fluctuations correlation occurs? External charges are screened on what length scales? These static observables are also physical quantities eminently suited for lattice experiments' measurements.

The usual perturbation theory has a severe infrared problem at the higher orders in coupling~\cite{Linde:1980ts, Gross:1980br}, but the infrared problems are only connected with the static, soft gluons. It has led to the development of the dimensional reduction technique ~\cite{Ginsparg:1980ef, Appelquist:1981vg}, which reduces the non-perturbative infrared behavior to a more straightforward effective theory for the soft modes only, enabling the precise estimation of the observables in the first order of expansion in perturbation theory. However, the systematic expansion in the perturbation theory for arbitrary orders remains unknown. This way, the terms involving dynamical quarks can be computed perturbatively in four dimensions. At the same time, the computationally costly non-perturbative methods can be used for a more straightforward three-dimensional theory. Naturally, the QCD coupling is significant at moderate temperatures, and higher-order corrections cannot be disregarded.

The dimensional reduction has been extensively used in QCD and electroweak theory to study thermodynamics ~\cite{Appelquist:1981vg, Kajantie:1995dw, Braaten:1995jr, Kajantie:2002wa, Blaizot:2003iq, Vuorinen:2003fs, Ipp:2006ij, Gynther:2005dj} and different gluonic correlators~\cite{Kajantie:1997tt, Laine:1997nq, Hart:2000ha, Hart:1999dj, Cucchieri:2001tw} in the bosonic sector. The fermionic modes, on the other hand, only affect the dimensionally reduced theory's parameters, which are generally integrated out from the effective theory. However, several intriguing observables built of quark fields are sensitive to infrared physics and, as a result, require some resummations, elegantly arranged as successive effective theories. The initial attempts to include this class of correlators in theory with reduced dimensions go back almost 30 years~\cite{Hansson:1991kb, Koch:1992nx}. These works, however, did not consistently include every term in a fixed order. In ref.~\cite{Huang:1995tz}, it was demonstrated that the dimensionally reduced theory could be expressed in nonrelativistic quarks, and the accurate power counting of various operators was established. The other motivation for studying the observables built of quark fields, i.e., ``mesonic and baryonic" observables, is related to their more closely signs with experimental signatures. In recent years, significant work has been done on one class of these observables, namely the various quark number susceptibilities (see ref.~\cite{Gottlieb:1987ac, Blaizot:2001vr, Gavai:2002kq, Gavai:2002jt, Chakraborty:2001kx, Chakraborty:2003uw, Vuorinen:2002ue, Andersen:2012wr, Haque:2013qta, Haque:2018eph}). 

The other class of observables, more closely related to IR physics and more sensitive toward it, is the correlation lengths of the mesonic operators, which are gauge-invariant quark bilinears constructed from the light quark flavors~\cite{Detar:1987kae}. The quark bilinears' two-point spatial correlation functions can determine the correlation lengths. Generally, these correlation functions are dominated by the screening masses at large distances, defined as the inverse of the correlation lengths. These screening masses give information about the quark-gluon plasma (QGP) response when a meson is included in the system. A large number of studies have been done in the literature to study the meson correlation function using perturbative methods as well as using non-perturbative lattice computations. On the perturbative side, the mesonic correlation functions using hard thermal loop approximation (HTL) have been studied in refs.~\cite{Karsch:2000gi, Alberico:2004we}. Additionally, the meson correlation function at finite momentum in QCD plasma has been studied in ref.~\cite{Alberico:2006wc}. Using the perturbative QCD, the leading order result $(2\pi T)$ for the meson screening mass is obtained in ref.~\cite{Eletsky:1988an, Florkowski:1993bq}. The next-to-leading (NLO) correction for meson and gluon screening masses has been calculated using an effective theory at zero chemical potential in ref.~\cite{Laine:2003bd, Laine:2009dh}. The connection between thermal screening masses and real-time rates has been explored in ref.~\cite{Brandt:2014uda} in the spectral representation. The NLO correction of meson screening masses at finite chemical potential has also been extended in ref.~\cite{Vepsalainen:2007ke}. 

On the lattice side, the temporal and spatial hadronic correlation function in QCD plasma has been studied in ref.~\cite{Born:1991zz, Boyd:1994np,QCD-TARO:2000hup}. Also, screening masses in purely SU(2) and SU(3) gauge theories using lattice QCD is studied in ref.~\cite{Datta:1998eb, Datta:1999yu, Cheng:2010fe, Gavai:2001ie}. Recently, mesonic screening masses have been studied for the first time in a large range covering the temperature from $\sim 1$ GeV to $\sim 160$ GeV using lattice QCD in ref.~\cite{DallaBrida:2021ddx}. Also, the recent lattice QCD results for the meson screening masses in $(2+1)$ flavor QCD in the temperature range from $0.14 \hspace{1mm} \text{GeV} \leq T \leq 2.7 \hspace{1mm}\text{GeV} $ is presented in ref.~\cite{Bazavov:2019www} and for the recent study of mesonic screening mass at finite chemical potential using lattice QCD see ref.~\cite{Thakkar:2022frk, Pushkina:2004wa}. In the high-temperature limit, the perturbative result of meson screening masses calculated in ref.~\cite{Laine:2003bd} is comparable with the lattice results. However, no such analytic calculation in the literature can explain the lattice data for  $T \leq 2$ GeV in the low-temperature regime. In this article, we have tried to overcome this gap by using a non-perturbative scheme offered by Gribov quantization.

Since QCD's infrared (IR) region is strongly coupled, the conventional resummed perturbative approach will not give appropriate results~\cite{Andersen:2004fp}. One of the effective ways to handle the IR region is to consider the Gribov-Zwanziger approach~\cite{Gribov:1977wm, Zwanziger:1989mf} (For recent reviews on Gribov-Zwanziger approach, see~\cite{Dokshitzer:2004ie, Vandersickel:2012tz} and some recent works on extended Gribov-Zwanziger approach follow~\cite{Dudal:2008sp, Capri:2016aqq, Dudal:2017kxb, Gotsman:2020ryd, Gotsman:2020mpg, Justo:2022vwa, Gracey:2010cg}) and the references therein. In these extended Gribov-Zwanziger models, first introduced in ref.~\cite{Dudal:2008sp}, the effect of local composite operators, which consist of a mass term, has been explored, and it has been found that the inclusion of the mass term in the propagator gives results which match very well with the lattice simulations results in the infrared domain. Also in ref.~\cite{Tissier:2010ts} the authors have calculated the gluon and ghost propagators to study the IR physics at zero temperature by including the mass term in usual Faddev Popov action, which are in excellent agreement with lattice results. Although the infrared domain of QCD is non-perturbative, once the Gribov copies effects are introduced, QCD can be treated perturbatively (or semi-perturbatively) in the infrared. Also, it has been observed in Monte-Carlo simulations that the Yang-Mills coupling constant in the infrared is compatible with a perturbative expansion once the mass effects in the propagator have been included. For more details on this, see the recent review in ref.~\cite{Pelaez:2021tpq}  
In recent years the general Gribov approach, without any mass term in the propagator, has been utilized to improve the thermodynamic quantities in QCD by evaluating free energy ~\cite{Fukushima:2013xsa}. Also, in kinetic theory, the transport coefficients have been explored using this scheme in ref.~\cite{Florkowski:2015dmm, Florkowski:2015rua,Jaiswal:2020qmj}. In this approach, the gluon propagator modifies its form, and a new (chromo)magnetic scale enters the theory through the mass parameter. Using the modified gluon propagator in Gribov action, quark dispersion relation~\cite{Su:2014rma}, the dilepton production rate along with quark number susceptability~\cite{Bandyopadhyay:2015wua} and electromagnetic debye mass~\cite{Bandyopadhyay:2023yjp} have been calculated, which sheds light on some of the new physical insights of the said observables. Recently, this approach has been applied to heavy quark phenomenology to study the heavy quark potential~\cite{Bandyopadhyay:2023yjp, Wu:2022nbv, Debnath:2023dhs} and heavy quark diffusion coefficient~\cite{Madni:2022bea}.

The following is how the paper is set up. We shall establish necessary notations and review some of the mesonic correlators' well-known characteristics at high temperatures in section~\ref{Section_2}. Section \ref{Section_3} covers how a dimensionally reduced effective field theory~\cite{Huang:1995tz} can describe high-temperature QCD in general while incorporating the mesonic correlators along the lines of work~\cite{Laine:2003bd}. 
In section \ref{Section_4}, we will 
do the required matching computation to fix the unknown parameter in the effective lagrangian by matching the dispersion relation of QCD and NRQCD$_{3}$, using Gribov formalism. Section \ref{Section_5} evaluates the dynamics of the effective theory with Gribov's gluon propagator inclusion. Results of the screening masses for the quenched QCD case and $N_f=3$ are compared with the recent lattice data in section \ref{Section_6}. We summarize the work in the last section, \ref{Section_7}. 
\section{ High-temperature static correlators in QCD}\label{Section_2}	
The Euclidean Lagrangian of the quarks at finite temperature QCD is given by
\begin{equation}\label{quark_lag.}
	\mathcal{L}_{\mathrm{E}}^{\mathrm{Q}}=\bar{\psi}\left(\gamma_\mu \mathcal{D}_\mu+M\right) \psi,
\end{equation}
Here, the covariant derivative ($\mathcal{D}_{\mu}$) of the fermionic field is given by
\begin{equation}
	\mathcal{D}_\mu \psi \equiv \partial_\mu \psi-i g A_\mu^a T^a \psi
\end{equation}
where $T^{a}$ are the generators defined in fundamental represenation of group $SU(N_c)$, which are hermitian in nature. The quark field $\psi$ is an $N_f$-component vector in flavor space. For the sake of simplicity, we will assume M to be a diagonal and degenerate matrix, $\text{M} = \text{diag}(m,\cdots m),$ and we will take the assumption of $m=0$. We may use quark fields to define bilinear objects with varying spin and flavor structures. We are interested in the operators having the form,
\begin{equation}\label{opera_form}
	\mathfrak{O}^{a} = \bar{\psi} f^{a}\Gamma\psi,
\end{equation}
where $\Gamma$ can take the values $\{1,\gamma_{5},\gamma_{\mu},\gamma_{\mu}\gamma_{5}\}$ for the different channels, namely scalar, pseudoscalar, vector, and pseudo vector $S^{a}, P^{a}, V_{\mu}^{a}$ and $A_{\mu}^{a}$ respectively. The traceless matrices $f^{a}$ along with the identity matrix $f^{s}$ provides the flavor basis,
\begin{eqnarray}
	f^{a}\equiv \{f^{s},f^{n}\} , \ f^{s} \equiv \mathbf{1}_{{N_f}*N_{f}}, \  \text{Tr}[f^{a} f^{b}] = \frac{1}{2}\delta^{ab} ,
\end{eqnarray}
with $a,b = 1,2,\cdots N^{2}_{f} - 1$. For the operators defined in eq.~\eqref{opera_form}, we will focus on the correlators having the structure as
\begin{equation}\label{corre_def}
	\mathcal{P}_{\mathbf{q}}\left[\mathfrak{O}^a, \mathfrak{O}^b\right] \equiv \int_0^{\frac{1}{T}} \mathrm{d} \tau \int \mathrm{d}^3 x \hspace{1mm} e^{i \mathbf{q} \cdot \mathbf{x}}\left\langle \mathfrak{O}^a(\tau, \mathbf{x}) \mathfrak{O}^b(0,0)\right\rangle,
\end{equation}
or, in position space,
\begin{equation}\label{Corr_in_position}
	\mathcal{P}_{\mathbf{x}}\left[\mathfrak{O}^a, \mathfrak{O}^b\right] \equiv \int_0^{1 / T} \mathrm{~d} \tau\left\langle \mathfrak{O}^a(\tau, \mathbf{x}) \mathfrak{O}^b(0,0)\right\rangle
\end{equation}

Using the rotational invariance, one can choose the correlation measurement direction as $z$ in eq.~\eqref{corre_def}. With this choice, one can take the average over the $x_{1} x_{2} $ surface, giving a correlator that is easier to handle
\begin{eqnarray}\label{corre_in_z}
	\mathcal{P}_z\left[\mathfrak{O}^a, \mathfrak{O}^b\right]&=&\int \mathrm{d}^2 \mathbf{x}_{\perp} \mathcal{P}_{\left(\mathbf{x}_{\perp}, z\right)}\left[\mathfrak{O}^a, \mathfrak{O}^b\right]
\end{eqnarray}
where $\mathcal{P}\left(\mathbf{x}_{\perp}, z\right)$ is given in eq.~\eqref{Corr_in_position}. The spatially separated two-point correlation function mentioned in eq.~\eqref{corre_in_z} defines the screening masses as 
\begin{equation}\label{screeM_def}
	m_{z}=-\lim _{z \rightarrow \infty} \frac{d}{dz} \ln \left[\mathcal{P}_z\left[\mathfrak{O}^a, \mathfrak{O}^b\right]\right]
\end{equation}
The above eq.~\eqref{screeM_def} describes the correlation function's exponential falloff at large distances.
\begin{figure}
	\centering
	\subfloat[\label{fig1(a)}]{\includegraphics[scale=1.]{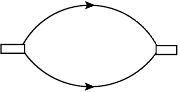}}
	\quad \quad \quad
	\subfloat[\label{fig1(b)}]{\includegraphics[scale=1]{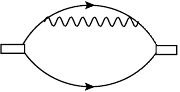}}
	\quad \quad \quad
	\subfloat[\label{fig1(c)}]{\includegraphics[scale=1.]{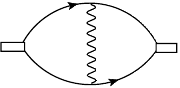}}
	\caption{The diagrams which contribute to meson correlation function: (a) free theory correlator (b) quark self-energy graph (c) interaction of quark and antiquark through gluon exchange.}%
	\label{fig:1}%
\end{figure}
Let us focus on the correlation function's behavior at high temperatures. Due to asymptotic freedom, the correlators with free fermions depicted in figure~(\ref{fig1(a)}) can be evaluated using perturbation theory. By using the dimensional regularisation technique, one would get the result for the diagram~(\ref{fig1(a)}) as
\begin{eqnarray}\label{corre_free}
	\mathcal{P}_{\boldsymbol{q}}\left[\mathfrak{O}^a, \mathfrak{O}^b\right] \!\!\!&=&\!\!\!\operatorname{Tr}\left[f^a f^b\right] N_c T \sum_{n=-\infty}^{\infty} \int \frac{\mathrm{d}^{3-2 \varepsilon} p}{(2 \pi)^{3-2 \varepsilon}}\nonumber \\
	&& \hspace{-2.7cm}\times  \frac{1}{\left[p_n^2+p^2\right]\left[p_n^2+(p+q)^2\right]} \operatorname{Tr}\left[(\slashed{p}+\slashed{q}) \Gamma^a \slashed{p} \Gamma^b\right]
\end{eqnarray}
Here, $N_c$ denotes number of colors, $\Gamma$ is the dirac matrix coming from the opeartor form $\mathfrak{O}^{a}$, $\slashed{p} \equiv \gamma_{\mu}p_{\mu}$ and $p_{n}$ refers to the fermionic Matsubara modes having the form, $p_{n} = 2\pi T (n+1/2)$. After doing the trivial Dirac algebra calculation, which gives some constant terms, the above correlation function in eq.~\eqref{corre_free} contains the function $A_{3\mathrm{d}}\left(2 p_n\right)$ which comes out to be
\begin{eqnarray}\label{A3d_func}
	A_{3\mathrm{d}}\left(2 p_n\right) 
	&=&\frac{i}{8 \pi q} \ln \frac{2 p_n-i q}{2 p_n+i q}.
\end{eqnarray}
From the eq.~\eqref{A3d_func}, it is clear that the singularity of the function $A_{3d}(2p_{0})$ appears at the point $2p_{n}$. Thus, the correlator dominates at large distances for the zeroth Matsubara frequencies defined as $\pm p_{0} = \pm \pi T$. Now, this frequency excitation occurs in the 3-dimensional theory. One can also consider this three-dimensional theory as $(2+1)$-dimensional and consider the time direction as the one in which the correlation is assessed. In this case, screening mass is equivalent to determining the screening states of a pair of on-shell heavy quarks, each with a mass of ``$p_0$''. 

\section{Effective Theory}\label{Section_3}
We want to determine the next-to-leading order (NLO) correction to the mesons screening mass using the non-perturbative effects. To go beyond the leading order, we must consider the diagrams shown in fig.~(\ref{fig1(b)}) and ~(\ref{fig1(c)}). Nevertheless, as we know, in finite temperature QCD, it will not be sufficient to calculate only these two diagrams. In principle, we can have a large number of diagrams that can behave in the same order as the two depicted in fig.~(\ref{fig1(b)}) and ~(\ref{fig1(c)}) do. 
In other words, to find the bound state (mesons) poles, one needs to do the entire sum of ladder diagrams since we can not obtain bound states by considering a finite number of Feynman diagrams. In our approach, we will use the diagrams above to find the interaction potential, which will be further used in the Schr\"odinger equation to find the non-relativistic bound states. This approach is equivalent to summing all the ladder diagrams in the non-relativistic limit.

To calculate the first-order correction to the free quarks propagator in a consistent way,  it is necessary to do the resummation of the diagrams with an arbitrary number of zero modes of the low-momentum exchanged gluon between quarks. The effective field theory approach provides a convenient way to perform such resummations. Only the diagram containing one-gluon exchanged would contribute to order $g^{2}$ depicted in fig.~(\ref{fig1(c)}). As shown in fig.~(\ref{fig1(c)}) for non-zero gluon modes, the gluon exchange diagrams do not need to be computed because this diagram alone cannot alter the pole location or the screening mass. These diagrams would contribute to the overall normalizing factor of the correlator under consideration. 

Conversely, in the bosonic section, the dynamics of the gluonic zero modes after the dimensional reduction technique describe the effective three-dimensional gauge theory~\cite{Appelquist:1981vg}, known as Electrostatic QCD(EQCD). The action takes the form as  
\begin{eqnarray}\label{EQCD_action}
\!\!\!	\mathcal{S}_{\mathrm{EQCD}}\!\!\!&=&\!\!\!\frac{1}{g_{\mathrm{E}}^2} \int d^3 x\left\{\frac{1}{2} \operatorname{Tr}\left[G_{i j} G_{i j}\right]+\operatorname{Tr}\left[\left(\mathcal{D}_j A_0\right) \right.\right.\nonumber\\
	&\times&\left.\left(\mathcal{D}_j A_0\right)\right] +\left.m_{\mathrm{E}}^2 \operatorname{Tr}\left[A_0^2\right]\right\}+\ldots,
\end{eqnarray}
where the dots refers to higher dimensional operators~\cite{Laine:2016hma} and $i =1,2,3$ , $G_{ij} = ig_{E}^{-1}[\mathcal{D}_{i},\mathcal{D}_{j}]$ , $\mathcal{D}_{i} = \partial_{i} - ig_{E}A_{i}$ , $g_{E}^{2} = g^{2}T$. The temporal component of gauge field $A_{0}$ acts as a scalar field with mass $m_{E}$. 
However, we are interested in the processes which occur at $\mathcal{O}(g_{E}^{2})$; then one can integrate out the scalar field from the theory in eq.~\eqref{EQCD_action}. After integrating the temporal part, we are left with the effective theory known as Magnetostatic QCD (MQCD), which is given by 
\begin{equation}\label{MQCD}
	\mathcal{S}_{\mathrm{MQCD}}=\frac{1}{g_{\mathrm{E}}^2} \int d^3 x\left\{\frac{1}{2} \operatorname{Tr}\left[G_{i j} G_{i j}\right]\right\}+\ldots
\end{equation}
This three-dimensional theory has non-perturbative dynamics; thus, it must be handled completely in a non-perturbative way~\cite{Linde:1980ts}. 

\subsection{Tree-level NRQCD$_{3}$}\label{Section_3.1}
In the fermionic sector, we will consider only the lowest fermionic modes $p_{0} \equiv \pi T$ because the other modes will not contribute much to the correlator at large distances as seen in eq.~\eqref{A3d_func}. The Euclidean quark lagrangian for the field $\psi(x)$ having Matsubara mode $\omega_{n}$ reads as 
\begin{equation}
	\mathcal{L}_{\mathrm{Q}}=\bar{\psi}\left[i \gamma_0 \omega_n-i g \gamma_0 A_0+\gamma_j \mathcal{D}_j+\gamma_3 \mathcal{D}_3\right] \psi
\end{equation}
where  $j=1,2$ and $A_{0}$ is the zero gluonic mode interacting with $\psi(x)$. 
We considered the Euclidean Dirac matrices a different form than the standard one. In this representation, we have
\begin{equation*}
	\gamma_0=\left(\begin{array}{cc}
		0 & 1_2 \\
		1_2 & 0
	\end{array}\right), \hspace{0.1cm} \gamma_i=\left(\begin{array}{cc}
		\epsilon_{i j} \sigma_j & 0 \\
		0 & -\epsilon_{i j} \sigma_j
	\end{array}\right), 
\end{equation*}
\begin{equation}	
	 \gamma_3=\left(\begin{array}{cc}
		0 & -i \\
		i & 0
	\end{array}\right),
\end{equation}
As the quarks are very heavy, the quark fields can be considered static fields, and the Dirac spinors can be rewritten in terms of $\chi$ and $\phi$
\begin{equation}
	\psi=\left(\begin{array}{l}
		\chi \\
		\phi
	\end{array}\right)
\end{equation}
where $\chi$, $\phi$ are the two-component spinor objects that led to the lagrangian 
\begin{eqnarray}
	\mathcal{L}_{\mathrm{Q}}&=&i \chi^{\dagger}\left(p_0-g A_0+\mathcal{D}_3\right) \chi -\chi^{\dagger} \epsilon_{i j} \mathcal{D}_i \sigma_j \phi \nonumber \\
	&+&\phi^{\dagger} \epsilon_{i j} \mathcal{D}_i \sigma_j \chi +i \phi^{\dagger}\left(p_0-g A_0-\mathcal{D}_3\right) \phi
\end{eqnarray}
where $\epsilon_{i j}$ is antisymmetric two rank tensor and $\epsilon_{12}=+1$. The quarks in the considered correlators are almost ``on-shell," and for the free theory, the on-shell point is given by $p_{0}^{2}+p^{2} = 0$, i.e., $p_{3} = \pm ip_{0}$. At the tree level, quarks with a fixed Matsubara frequency $\omega_{n}$ interact with zero gluonic modes only; thus, it is expected that quarks' off-shellness is related to the gluonic momentum scale as $|p_{3}\pm i p_{0}|\lesssim gT$. Now, after expanding around a state consisting of a free pair of quark-antiquark, one of the components is heavy, and another is light compared to the dynamical scale, viz. $gT$, $g^{2}T$, one can solve the equation of motion for the heavy component and the light mode. After doing an expansion in powers of $ 1/p_{0}$, the effective action for the fermionic mode can be written as 
\begin{eqnarray}\label{tree_level_action}
	\mathcal{S}_\psi^{\text {eff}} \!\!&=&\!\! \int d^3 x \{ i \chi^{\dagger}\left[p_{0}-g_{\mathrm{E}} A_0+\mathcal{D}_3-\frac{1}{2 p_{0}}\left(\mathcal{D}_k^2+\frac{g_{\mathrm{E}}}{4 i}\nonumber\right.\right.\\
	&\times& \left.\left. \left[\sigma_k, \sigma_l\right] G_{k l}\right)\right]\chi + i \phi^{\dagger}\left[p_{0}-g_{\mathrm{E}} A_0-\mathcal{D}_3 -\frac{1}{2 p_{0}}\right.\nonumber\\
	&\times& \left.\left.\left(\mathcal{D}_k^2+\frac{g_{\mathrm{E}}}{4 i}\left[\sigma_k, \sigma_l\right] G_{k l}\right)\right] \phi\right\}+ \mathcal{O}\bigg(\frac{1}{p_{0}^{2}}\bigg)
\end{eqnarray}

\subsection{{Power Counting Arguments}}\label{Section_3.2}
We must consider the quantum corrections subjected to the parameters in eq.~\eqref{tree_level_action} to ascertain the radiative corrections. One must take into account all additional operators that symmetries might permit. Setting up a power counting for the likely operators is crucial. 

In the NRQCD$_{3}$ side, quarks have momentum scale as $|\textbf{p}_{\perp}| \lesssim gT$ and their off-shellness $\Delta p_{3} = p_{3}+ip_{0}$ is of the same order. By requiring the action mentioned in eq.~\eqref{tree_level_action} to be of the order of unity, we have
\begin{equation}
	\begin{aligned}
		& \int \mathrm{d} z \mathrm{~d}^2 \mathrm{x}_{\perp} \hspace{1mm} \chi^{\dagger} \partial_3 \chi \sim 1, \Rightarrow \chi \sim 1 /\left|\mathrm{x}_{\perp}\right| \sim g T \\
		& \int \mathrm{d} z \mathrm{~d}^2 \mathrm{x}_{\perp} \hspace{1mm} A \partial_3^2 A \sim 1 \Rightarrow A  \sim g^{1 / 2} T^{1 / 2}
	\end{aligned}
\end{equation}
On-shell relativistic gluons have the same energy and momentum order viz., $p_{3} \sim \textbf{p}_{\perp}$. On the other hand, as the considered quarks are nonrelativistic,  their kinetic energy is directly proportional to momentum squared in $(2+1)$ dimensional theory. For a nearly on-shell quark with transverse momentum $|\textbf{p}_{\perp}| \lesssim gT$ in NRQCD$_{3}$ , the off-shellness in longitudinal momentum becomes
$		\Delta p_3 \sim \mathbf{p}_{\perp}^2 / p_0 \sim g^2 T$ and this gives $\partial_{3}\sim g^2T$ that acts on quarks.


Similar power counting arguments demonstrate that $g_{E}A \sim g^{3/2}T$, which means that this term will be of higher order compared to the derivative term $\partial_{i}$ present in the transversal covariant derivative. As $\partial_{\perp}^{2}$ is already of $\mathcal{O}(g^{2})$, one can left out the trasnverse gluons. Thus the only parameter which needs to be matched beyond tree-level is the zero point energy $p_{0}\equiv M$. Thus to find out the $\mathcal{O}(g^{2})$ corrections to meson correlation lengths, the following lagrangian is sufficient to use  
\begin{eqnarray}\label{effective_lag.}
	\mathcal{L}_{\text {eff }}^{\mathrm{\psi}}&=&i \chi^{\dagger}\left(M-g_{\mathrm{E}} A_0+\mathcal{D}_3-\frac{\nabla_{\perp}^2}{2 p_0}\right) \chi\nonumber \\
	&+&i \phi^{\dagger}\left(M-g_{\mathrm{E}} A_0-\mathcal{D}_3-\frac{\nabla_{\perp}^2}{2 p_0}\right) \phi ,
\end{eqnarray}
where $\mathcal{D}_{3}= \partial_{3} - ig_{E} A_{3}$. Note that, here $A_{0}$ and $A_{3}$ play an important dynamical scale role. On the other hand, transverse gluons $A_{1}$, $A_{2}$ can be ignored as long as one is interested in the energy shift of $\mathcal{O}(g^{2}T)$. This means that gluons will not transfer transverse momentum to quarks in this order. To be consistent at $\mathcal{O}(g^{2}T)$, we should replace, energy i.e., $p_{0}$ of tree-level effective Lagrangian by a matching coefficient $M = p_{0} + \mathcal{O}(g^{2}T)$, which we will determine in the next section.   

\section{Matching Conditions from QCD to NRQCD$_{3}$, with Gribov}\label{Section_4}
We will determine the one-loop correction to the variable $M$ by matching the Green function, calculated on the QCD side and NRQCD$_{3}$ side, using Gribov formalism. Here, the matching will be done by finding the finite temperature Euclidean dispersion relation so that we do not need to worry about the overall normalization factors arising from the fields. This computation produces gauge-invariant results in both sectors, and we will do the matching order by order in $1/p_{0}$ on the NRQCD$_{3}$ side so that dimensional regularisation copes up with the power counting inputs as done in ref.~\cite{Manohar:1997qy}. 

In the conventional quantization of QCD, the gauge condition is not ideal, as proposed by Faddeev and Popov. Thus, there remains a residual gauge discrepancy in the infrared region, which is pointed out by Gribov and known as the Gribov copy problem~\cite{Gribov:1977wm}. Gribov proposed that the functional integral should be constrained to the first Gribov region, which collects gauge fields without zero modes for the Faddeev-Popov operator. 
The expression for the Gluon propagator with Gribov quantization in the general covariant gauge is written as
\begin{equation}\label{Gribov_prop}
	D_{\mu\nu}^{ab}(P)=\delta^{a b} \left(\delta_{\mu \nu}-(1-\xi)\frac{P_\mu P_\nu}{P^2}\right)\frac{P^2}{P^4+\gamma_{\mathrm{G}}^4}
\end{equation}
where $\xi = 0,1$ corresponds to Landau and Feynman gauges. On the QCD side, using the Gribov propagator mentioned in eq.~\eqref{Gribov_prop}, the quark propagator's inverse, defined as $S^{-1}(P)$, is given by,
\begin{eqnarray}\label{Disp_QCD}
	\hspace{-0cm}	-iS^{-1}(P)&=& \slashed{P} -\left. g^{2} C_{F} \SumInt_{Q} \frac{\gamma_{\mu} (\slashed {P}-\slashed{Q}) \gamma_{\mu}}{(P-Q)_{f}^{2}}\left(\frac{Q^{2}}{Q^{4}+\gamma_{G}^{4}}\right)_{b} \right.\nonumber \\
	&&\hspace{-2.2cm}+ g^{2} C_{F}\left.\SumInt_{Q} \frac{\slashed{Q} (\slashed {P}-\slashed{Q}) \slashed{Q}}{Q^{2}(P-Q)_{f}^{2}}\left(\frac{Q^{2}}{Q^{4}+\gamma_{G}^{4}}-\frac{\xi\hspace{1mm} Q^{2}}{Q^{4}+\gamma_{G}^{4}}\right)_{b}\right.
\end{eqnarray}
where the gluonic four-momentum in Euclidean space-time reads $Q=\left(q_{0},q\right)$, with $q_0=2 n \pi T$, and $N_c$ represents the number of colors. This calculation uses a dimensional regularization technique with $\overline{\mathrm{MS}}$ renormalization scheme. In this technique, the sum-integral is given as
\begin{equation}
	\SumInt_{Q} \equiv\left(\frac{e^{\gamma_E} \Lambda^2}{4 \pi}\right)^\varepsilon T \sum_{q_0=2 n \pi T} \int \frac{d^{d-1} q}{(2 \pi)^{d-1}},
\end{equation}
with $d=4-2 \varepsilon$ as the space-time dimensions. Also, $(\cdots)_{f} , (\cdots)_{b}$ refers to fermionic and bosonic Matsubara modes respectively, and $C_{F} = (N_{c}^{2}-1)/2N_{c}$. 
As we are calculating the $\mathcal{O}(g^{2})$ corrections to $\Sigma(P)$, one can utilize the free theory constraints namely $P^{2}=0$ and $\slashed{P} u(P) =0$ to handle the terms which are in proportion to $P^{2}$ and $\slashed{P}$. Thus the longitudinal gauge-dependent part then vanishes, leaving only the transverse terms of Gribov's gluon propagator to contribute to the calculation. The outcome is irrespective of the gauge component $\xi$ as expected.

Because of the rotational invariance of the remaining terms in eq.~\eqref{Disp_QCD}, we can do the matching of the Green's function at some particular momentum and set $p_{\perp} = 0$. After the multiplication of $\gamma_{0}$ from the left to eq.~\eqref{Disp_QCD}, the expression becomes block-diagonal. Thus we can focus on the particular term, say, $[\gamma_{0}S^{-1}(P)]_{11}$, which is given by 
	\begin{eqnarray}\label{gamma0_Sigma}
	\hspace{-.2cm}	-i	\left[\gamma_0 S^{-1}(P)\right]_{11}\!\!\!&=&\!\!\!\! p_0+i p_3 -g^2 C_F \SumInt_Q \left(\frac{Q^{2}}{Q^{4}+\gamma_{G}^{4}}\right)_{b} \nonumber\\
		&\times& \frac{2(-p_{0}-ip_{3}+q_{0}+iq_{3})}{(P-Q)_f^2}
	\end{eqnarray}
In eq.~\eqref{gamma0_Sigma}, we have the contribution coming from the transverse gluons, i.e., $A_{1}$ and $A_{2}$ component only. In contrast, the gauge field components $A_{0}$ and $A_{3}$ vanish altogether at the free theory pole position $p_{0}=-ip_{3}$ as they occur with opposite signs. Thus the only integral which we need to evaluate to find the $\mathcal{O}(g^{2})$ correction is 
	\begin{eqnarray}
		I	&=&\left.\SumInt_{Q} \frac{2\left(q_{0}+i q_{3}\right)}{(P-Q)_{f}^{2}}\left(\frac{Q^2}{Q^4+\gamma_{G}^4}\right)_{b}\right|_{p_{0}=-i p_{3}}\nonumber\\
		&=& I_1 +  I_2.
	\end{eqnarray}
	Now, the first integral  $I_1$ becomes
	\begin{eqnarray}\label{int_I1}
		I_1 = - \hspace{1mm} \frac{1}{p_{0}} \int_{0}^{\infty} \frac{q^{2}dq}{(2\pi)^{2}}\bigg[\frac{n^+}{E_+} +	 \frac{n^-}{E_-} \bigg].
	\end{eqnarray} 
	Here,  $n^\pm$ represents the Bose-Einstein (B.E) distribution function with $ E_\pm=\sqrt{q^2\pm i\gamma_{G}^2} $. Additionally, the second integral $I_{2}$ takes the form as 
	
	\begin{eqnarray}\label{int_I2}
		I_2 
		& =&\frac{1}{p_{0}}\bigg[ \frac{-T^{2}}{24} + X \bigg],
	\end{eqnarray}
	where $X$ is given by
	
	\begin{eqnarray}\label{def._X}
	\hspace{-0cm}	X &=& \frac{\gamma_{G}^{4}}{T^2} \int \frac{ d^3q}{(2\pi)^3}\bigg[\left(\frac{\tilde{n}+n^{-}}{i\pi-E+E_{-}} - \frac{\tilde{n}+n^{+}}{i\pi-E+E_{+}}\right)\nonumber\\
		&+&  \left(\frac{\tilde{n}+n^{+}}{i\pi+E-E_{+}} - \frac{\tilde{n}+n^{-}}{i\pi+E-E_{-}}\right)\bigg]\nonumber \\
	&\times&	\frac{1}{8E E_{+} E_{-}}\frac{1}{\left(E_{+}-E_{-}\right)} 
	\end{eqnarray}
	where $\tilde{n}$ is the Fermi-Dirac (F.D) distribution function with energy $E = q-p_3(1+\cos\theta)$. 
	Thus, the Euclidean dispersion relation on the QCD side using the Gribov propagator becomes
	\begin{equation}
		p_{3} \approx i\bigg[p_{0}-g^{2}C_{F}(I_{1}+I_{2})\bigg]
	\end{equation}
	On the NRQCD$_{3}$ side, one can extract the Feynman rules from the lagrangian in eq.~\eqref{effective_lag.}. The one-loop contribution to the quark self-energy vanishes on NRQCD$_{3}$ side because the two gauge field components, namely $A_{0}$ and $A_{3}$, are the same and come with opposite signs as is the case in QCD side. Thus the inverse propagator in NRQCD$_{3}$ becomes
	
\begin{eqnarray}
\hspace{-.3cm}	-i S^{\prime{-1}}(P)\hspace{-.1cm}&=&\hspace{-.2cm}M+i p_3-g_{\mathrm{E}}^2 C_F \int \frac{\mathrm{d}^{3-2 \epsilon} q}{(2 \pi)^{3-2 \epsilon}} \frac{1}{M+i p_3-i q_3}\nonumber \\
	&\times&	\left[\left(\frac{Q^{2}}{Q^{4}+\gamma_{G}^{4}}\right)_{A_{0}}-\left(\frac{Q^{2}}{Q^{4}+\gamma_{G}^{4}}\right)_{A_{3}}\right]
\end{eqnarray}
So, the pole location is simply $p_{3} = i M$ on NRQCD$_{3}$ side. Now, after doing the matching, we will get 
\begin{equation}\label{def._M}
	M = p_{0}-g^{2}C_{F}(I_{1}+I_{2})
\end{equation}
The integrals $I_{1}$ and $I_{2}$ need to be evaluated numerically to find the matching parameter $M$. This matching accounts only for the hard gluons in the loop integral of figure~(\ref{fig1(b)}).
\section{Dynamics of effective theory}\label{Section_5}
After considering the hard gluons for figure~(\ref{fig1(b)}), we still need to take care of the soft gluons involved in figure~(\ref{fig1(b)}) and as well as fig.~(\ref{fig1(c)}), i.e., we need to solve the dynamics of the effective theory at $\mathcal{O}(g^{2}T)$. The dynamics will be studied by computing the correlators, as done in section \ref{Section_2}, by using the effective theory mentioned in eq.~\eqref{MQCD},~\eqref{effective_lag.}. 
The modified form of the correlation function of eq.~\eqref{corre_in_z} is
\begin{equation}
	\mathcal{P}_z\left[\mathfrak{O}^a, \mathfrak{O}^b\right] \sim \int \mathrm{d}^2 x_{\perp}\left\langle \mathfrak{O}^a\left(x_{\perp}, t\right) \mathfrak{O}^b\left(\mathbf{0}_{\perp}, 0\right)\right\rangle
\end{equation}
One may assume that the exchange of gluons between quarks and antiquarks occurs instantly in the nonrelativistic domain and may find out the static potential for the quark-antiquark pair by integrating the gauge fields. After obtaining the potential, we may use the usual Schr\"odinger equation to solve for the screening states and arrive at a solution. 

Thus we need to establish Green's function for the quantity of the form $\int \mathrm{d}^2 x_{\perp}\phi_u^*\left(\mathbf{x_{\perp}}, z\right) \Gamma_{u v} \chi_v\left(\mathbf{x_{\perp}}, z\right)$, where $\Gamma_{u v}$ is a constant term which consists of $2\times2$ matrix which does not affect the overall calculation if we drop out that term. To find out the static potential, one can use the trick of introducing a point-splitting in the correlator, which means putting the quark pairs at a particular distance $\mathbf{r}$ apart, which is finite, and then determining the Schr\"odinger equation obeyed by the correlator and set $\mathbf{r} \rightarrow 0 $ afterward. Thus the updated correlator is   
\begin{equation}
	\mathcal{P}\left(\mathbf{r}, z\right) \equiv \int_{\mathbf{R}}\left\langle\phi^{\dagger}\left(\mathbf{R}+\frac{\mathbf{r}}{2}, z\right)  \chi\left(\mathbf{R}-\frac{\mathbf{r}}{2}, z\right) \chi^{\dagger}(0) \phi(0)\right\rangle,
\end{equation}
For the Gribov's gluon propagator $D_{\mu \nu}^{a b}(P)$, we have
\begin{eqnarray}
\hspace{-.4cm}	\left\langle A_0^a(P) A_0^b(Q)\right\rangle&=&\delta^{a b}(2 \pi)^{4-2 \epsilon} \delta^{(4-2 \epsilon)}(P+Q) \frac{P^{2}}{P^{4}+\gamma_{G}^{4}} \nonumber\\
\hspace{-.4cm}	\left\langle A_i^a(P) A_j^b(Q)\right\rangle&=&\delta^{a b}(2 \pi)^{4-2 \epsilon} \delta^{(4-2 \epsilon)}(P+Q)\left[\frac{P^{2}}{P^{4}+\gamma_{G}^{4}}\right. \nonumber \\
\hspace{-1cm}	&\times& \left(\delta_{i j}-\frac{P_i P_j}{P^2}\right)\left. +\frac{P_i P_j}{P^2} \frac{\xi P^{2}}{P^{4}+\gamma_{G}^{4}}\right] .
\end{eqnarray}
In general, the equation of motion satisfied by Green's function, for large values of $z$, is of the form $(\partial_{z}-H)G(z)=c\hspace{0.5mm} \delta(z)$, where $H$ is hamiltonian of the system and $c$ denotes a constant value. Thus the equation of motion obeyed by the correlation functions at the tree level reads as  
\begin{equation}\label{tree-level_EOM}
	\left[\partial_z + 2 M-\frac{1}{p_0} \nabla_{\boldsymbol{r}}^2\right] \mathcal{P}^{(0)}(r, z) = 2 N_{c}\hspace{0.5mm} \delta(z) \hspace{0.5mm}\delta^{(2)}(\mathbf{r}),
\end{equation}
where the Hamiltonian can be read from the eq.~\eqref{effective_lag.}. For the one-loop diagrams shown in fig.~(\ref{fig1(b)}) and (\ref{fig1(c)}), the equation of motion is given by
\begin{eqnarray}\label{one-loop_EOM}
	&&	\hspace{-1cm}\left[\partial_z+2 M-\frac{1}{p_0} \nabla_{\boldsymbol{r}}^2\right] \mathcal{P}^{(1)}(\boldsymbol{r}, z)=-g_{\mathrm{E}}^2 C_F \mathcal{P}^{(0)}(\boldsymbol{r}, z) \nonumber \\
	&\times&	\mathcal{K}\left(\frac{1}{z p_0}, \frac{\nabla_{\boldsymbol{r}}}{p_0}, \frac{\gamma_{G}^{4}}{p_0^4}, \boldsymbol{r} p_0\right) 
\end{eqnarray}
where the kernel $\mathcal{K}$ is a dimensionless quantity that can be expanded to its first two arguments to find out the one-loop static potential, 
\begin{eqnarray}\label{static_poten.}
	\Phi(r) & \equiv & g_{\mathrm{E}}^2 C_F  \mathcal{K}\left(0,0, \frac{\gamma_{G}^{4}}{p_0^4}, r p_0\right)\nonumber\\ 
	&=&-2g_{\mathrm{E}}^2 C_F \int \frac{\mathrm{d}^{2} q}{(2 \pi)^{2}}
	\times	\frac{Q^{2}}{Q^{4}+\gamma_{G}^{4}}\ e^{i \boldsymbol{q} \cdot \boldsymbol{r}}
\end{eqnarray}
Now the eqs.~\eqref{tree-level_EOM},~\eqref{one-loop_EOM},~\eqref{static_poten.} can be clubbed together to get the final equation of motion upto one-loop order as
\begin{equation}
	\left[\partial_z+2 M-\frac{1}{p_0} \nabla_{\boldsymbol{r}}^2+\Phi(r)\right] \mathcal{P}(r, z) = 2 N_{c} \delta(z) \delta^{(2)}(\boldsymbol{r}) ,
\end{equation}
For the Gribov propagator, the expectation values of temporal gauge fields and spatial gauge fields are the same in the computation of the static potential. 
After solving the above integral in eq.~\eqref{static_poten.}, we get the final form of one-loop static potential as
\begin{equation}
	\Phi(r)=g_{\mathrm{E}}^2 \frac{C_F}{2 \pi}\left[\ln \frac{{\gamma_{G}} r}{2}+\gamma_E-K_0\left(\gamma_{G} r\right)\right] .
\end{equation}
where $K_{0}$ is the modified Bessel function whose asymptotic behavior is given by $K_{0}(y) = -\ln\frac{y}{2}-\gamma_{E}+\mathcal{O}(y)$. Now, this potential will determine the correlation length $\zeta^{-1}= m $, through
\begin{equation}\label{sch_eq.}
	\left[2 M-\frac{\nabla_{\boldsymbol{r}}^2}{p_0}+\Phi(\mathbf{r})\right] \Psi_0 = m \Psi_0 ,
\end{equation}
where $\Psi_{0}$ represents the ground state wave function. We will numerically solve this eq.~\eqref{sch_eq.} to find the screening mass. For the easiness of the problem, we can do the re-scaling as 
\begin{equation}\label{screening_mass_final}
	r \equiv \frac{r^{\prime}}{\gamma_{G}}, \quad m -2 M \equiv g_{\mathrm{E}}^2 \frac{C_F}{2 \pi} E_0
\end{equation}
In the polar coordinates, the Schr\"odinger equation~\eqref{sch_eq.} modifies as
\begin{equation}\label{sch_eq_polar}
	\hspace{-0.45cm}	\left[\frac{\mathrm{d}^2}{\mathrm{d} r^{\prime 2}}+\frac{1}{r^{\prime}} \frac{\mathrm{d}}{\mathrm{d} r^{\prime}}-\lambda\left(\ln \frac{r^{\prime}}{2}+\gamma_E-K_0(r^{\prime})- E_0\right)\right] \Psi_0=0,
\end{equation}
where
$
\lambda=p_0 g_{\mathrm{E}}^2 C_F/\left(2 \pi \gamma_{G}^{2}\right).
$
To find out the numerical solution of eq.~\eqref{sch_eq_polar}, we need to find out the wave function $\psi_{0}(r^{\prime})$ around the origin. Thus one finds that~\cite{Laine:2003bd}, 
\begin{equation}
\hspace{-.3cm}	\Psi_0(r^{\prime}) \approx \Psi_0(0)\left[1+\frac{1}{2} \lambda r^{\prime 2}\left(\ln \frac{r^{\prime}}{2}+\gamma_E-1-\frac{E_0}{2} \right)\right]. 
\end{equation}
$\Psi_{0}(0)$ is some finite number to make the wave function bounded. To find out the $E_{0}$, one can integrate the eq.~\eqref{sch_eq_polar} for a large value of $r^{\prime}$ and by requiring the square integrability condition. Contrary to the perturbative case~\cite{Laine:2003bd}, we find temperature dependent $E_{0}$ that saturates to a constant value at high temperature.
\section{Results and Discussion}\label{Section_6}
This section will outline our findings for the screening masses and contrast them with the most recent lattice data obtained in ref.~\cite{Bazavov:2019www}. The screening mass can be obtained from the eqs.~\eqref{def._M} and~\eqref{screening_mass_final} as 
\begin{eqnarray}\label{screening_mass_Final}
	m&= &2\pi T + g^{2}T \frac{C_{F}}{2\pi} \bigg(E_{0} - \frac{4\pi}{T}\big(I_{1}+I_{2}\big)\bigg)
\end{eqnarray}
In the calculation of screening mass, we have used the lattice-fitted running coupling obtained in ref.~\cite{Fukushima:2013xsa}, which is given by
\begin{equation}
	\alpha_s\left(T / T_c\right) \equiv \frac{g^2\left(T / T_c\right)}{4 \pi}=\frac{6 \pi}{11 N_{\mathrm{c}} \ln \left[a\left(T / T_c\right)\right]},\label{alpha}
\end{equation}
where $a = 1.43$ for the infrared (IR) case and $a=2.97$ for ultraviolet (UV) case respectively. To obtain the one loop coupling in eq.~\eqref{alpha}, the authors fitted the lattice data of running coupling extracted from the IR and UV behavior of heavy-quark free energy from ref.~\cite{Kaczmarek:2004gv}. In our calculation, since we are looking at the non-perturbative region, we use the infrared coupling case. 
\begin{figure}[b]
	\centering
	\captionsetup{width=0.45\textwidth}
	\includegraphics[width=8cm,height=9cm,keepaspectratio]{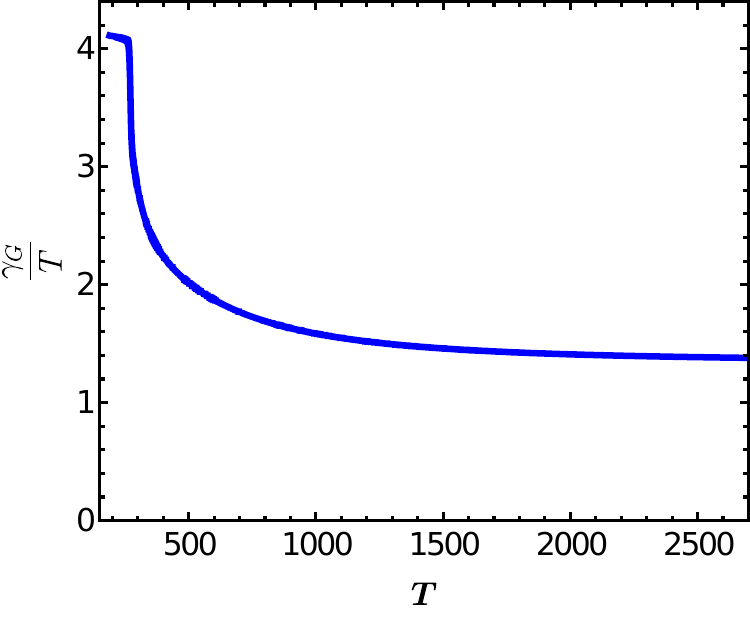}
	\caption{Temperature variation of scaled Gribov mass parameter obtained using lattice (thermodynamics) data.}
	\label{gammaG_vs_T}
\end{figure}
The integrals $I_{1}$ and $I_{2}$ consist of the Gribov parameter $\gamma_{G}$, which is often determined utilizing the one-loop or two-loop gap equation (for details, see~\cite{Gracey:2005cx}). Figure~\ref{gammaG_vs_T} shows the temperature variation of the scaled Gribov mass parameter $\gamma_{G}/T$ has been obtained after doing the matching with the lattice thermodynamics data, as obtained in ref.~\cite{Jaiswal:2020qmj}. For the values of $\gamma_{G}/T$ shown in figure~\ref{gammaG_vs_T}\, the integrals $I_{1}$ and $I_{2}$ have been evaluated numerically. The integral $I_{2}$ also contains the imaginary part, which eventually comes from the pole condition used to evaluate integral $X$ shown in eq.~\eqref{def._X}. However, since screening mass depends on the real value of parameter $M$ defined in eq.~\eqref{def._M}, by using eq.~\eqref{screening_mass_Final}, we plot the scaled screening mass $m/T$ with temperature in figure~\ref{screening_mass_vs_T} for quenched QCD $(N_{f}=0)$ case and for $N_{f} = 3$ case. We found a good agreement with the lattice data reported in ref.~\cite{Bazavov:2019www} and a good improvement over the perturbative results obtained in ref.~\cite{Laine:2003bd} in the low-temperature domain. The main outcome of fig.~\ref{screening_mass_vs_T} shows that screening mass significantly decreases from the free theory results in the low-temperature region. At the same time, in the high-temperature realm, it approaches the screening mass result obtained in ref.~\cite{Laine:2003bd}.
	\begin{figure}[tbh]
		\centering
		\captionsetup{width=0.45\textwidth}
		\includegraphics[width=8.5cm,height=6.5cm]{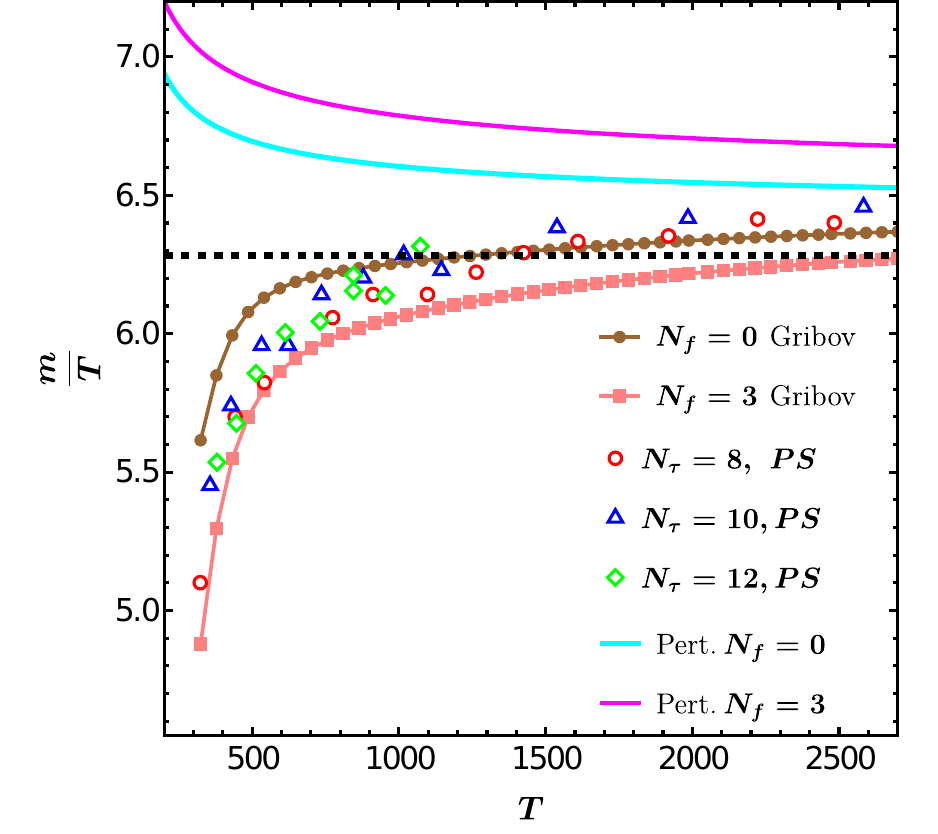}
		\caption{The temperature dependence of the scaled screening mass. The dashed line represents the free theory result from $(m =2\pi T)$. We compare the Gribov results for quenched and $(2+1)$ flavor QCD case with the perturbative and lattice results for various $N_{\tau}$. Here, $PS$ represents the pseudo scalar channel.}
		\label{screening_mass_vs_T}
	\end{figure}
\section{Summary}\label{Section_7}
In this work, we used the non-perturbative resummation using the Gribov quantization approach to study the mesonic screening masses. We began by examining the fundamental characteristics of the static meson correlators at high temperatures and saw that the zeroth fermionic Matsubara mode dominates the large distance correlator. Then we discussed the framework of the effective theory, namely NRQCD$_{3}$, in which we need to determine the parameters of the effective lagrangian to find out the non-perturbative NLO correction to mesonic screening mass. This matching is done by finding the Euclidean dispersion relation on the NRQCD$_{3}$ side and QCD side, using the Gribov approach. After this, we calculated the static quark-antiquark potential using Gribov's gluon propagator in the context of effective theory to understand the dynamics of the theory. This static potential determines the coefficient of exponential falloff by numerically solving the usual Schr\"odinger equation. We plotted the screening mass in the temperature ranges from$~300$ MeV to $2700$ MeV for $N_{f} = 0$ and $N_{f} = 3$, respectively. We compare our results with the recently obtained lattice data~\cite{Bazavov:2019www} and found a good agreement. Since our calculation does not respect the different channels like a scalar, vector, etc., an immediate extension can be done by adopting some strategy to include the different channels and different mesons in this framework. The other natural extension can be done by having the finite chemical potential in theory for a more general case study of mesonic screening masses which can give insights into the nature of the QCD phase diagram, which we will report soon.

\section*{Acknowledgements}
	The authors want to thank Mikko Laine and Michael Strickland for the valuable discussions. Sumit would like to acknowledge the hospitality of NISER, where most of the work is done. N.H. is supported in part by the SERB-Mathematical Research Impact Centric Support (MATRICS) under Grant No. MTR/2021/000939.

\appendix

\end{document}